\documentclass[11pt,letterpaper]{article}
\usepackage{latexsym}

\makeatletter
                                                                               
\def\icite{\@ifnextchar [{\@tempswatrue\@citey}{\@tempswafalse\@citey[]}}
\def\@citex[#1]#2{%
\if@filesw \immediate \write \@auxout {\string \citation {#2}}\fi
\@tempcntb\m@ne \let\@h@ld\relax \def\@citea{}%
\@cite{%
  \@for \@citeb:=#2\do {%
    \@ifundefined {b@\@citeb}%
      {\@h@ld\@citea\@tempcntb\m@ne{\bf ?}%
      \@warning {Citation `\@citeb ' on page \thepage \space undefined}}%
      {\@tempcnta\@tempcntb \advance\@tempcnta\@ne%
      \@tempcntb\number\csname b@\@citeb \endcsname \relax%
      \ifnum\@tempcnta=\@tempcntb 
        \ifx\@h@ld\relax%
          \edef \@h@ld{\@citea\csname b@\@citeb\endcsname}%
        \else%
          \edef\@h@ld{\ifmmode{-}\else--\fi\csname b@\@citeb\endcsname}%
        \fi%
      \else
        \@h@ld\@citea\csname b@\@citeb \endcsname%
        \let\@h@ld\relax%
      \fi}%
    \def\@citea{,\penalty\@highpenalty\,}%
  }\@h@ld
}{#1}}
\def\@citey[#1]#2{%
\if@filesw \immediate \write \@auxout {\string \citation {#2}}\fi
\@tempcntb\m@ne \let\@h@ld\relax \def\@citea{}%
\@icite{%
  \@for \@citeb:=#2\do {%
    \@ifundefined {b@\@citeb}%
      {\@h@ld\@citea\@tempcntb\m@ne{\bf ?}%
      \@warning {Citation `\@citeb ' on page \thepage \space undefined}}%
      {\@tempcnta\@tempcntb \advance\@tempcnta\@ne%
      \@tempcntb\number\csname b@\@citeb \endcsname \relax%
      \ifnum\@tempcnta=\@tempcntb 
        \ifx\@h@ld\relax%
          \edef \@h@ld{\@citea\csname b@\@citeb\endcsname}%
        \else%
          \edef\@h@ld{\ifmmode{-}\else--\fi\csname b@\@citeb\endcsname}%
        \fi%
      \else
        \@h@ld\@citea\csname b@\@citeb \endcsname%
        \let\@h@ld\relax%
      \fi}%
    \def\@citea{,\penalty\@highpenalty\,}%
  }\@h@ld
}{#1}}
\def\@cite#1#2{{$^{#1}$\if@tempswa , #2\fi }}
\def\@icite#1#2{{$#1$\if@tempswa , #2\fi }}

\gdef\@publabel{\hfil}
\gdef\@pubdate{\null}
\gdef\@pubnumber{\null}
\gdef\@author{\null}
\gdef\@title{\null}
\gdef\@abstract{\null}
\long\def\pubdate#1{\gdef\@pubdate{#1}}
\long\def\pubnumber#1{\gdef\@pubnumber{#1}}
\long\def\publabel#1{\gdef\@publabel{#1}}
\long\def\author#1{\gdef\@author{#1}}
\long\def\title#1{\gdef\@title{#1}}
\long\def\abstract#1{\gdef\@abstract{#1}}
\def\titlerelax{
\let\maketitle\relax
\let\settitleparameters\relax
\let\consolidatetitle\relax
\let\inittitlepage\relax
\let\finishtitlepage\relax
\let\titlepagecontents\relax
\let\multithanks\relax
\let\titlebaselines\relax
\let\@makepub\relax
\let\@maketitle\relax
\let\@makeauthor\relax
\let\@makeabstract\relax
\let\@maketitlenote\relax
\let\thanks\relax
\let\titlerelax\relax}

\def\titleclean
{\gdef\@titlenote{}
\gdef\@abstract{}
\gdef\@author{}
\gdef\@title{}
\gdef\@pubdate{}\gdef\@pubnumber{}\gdef\@publabel{}
\gdef\@dpublabel{}
}
\def\@makepub{\vbox to \z@{\hbox to \textwidth{\hfill
\@publabel \hfill
\llap{\parbox[t]{0.25\textwidth}{\raggedleft\@pubnumber}}}%
\vss}}
\def\@maketitle{\vskip 60pt \begin{center}
 {\LARGE \@title \par}
 \end{center}}
\def\@makeauthor{{%
\def\and{\smallskip {\normalsize \rm and\smallskip }}
\def\And{\medskip {\normalsize \rm and\\}\medskip}
\long\def\address##1{{\def\and{\\and\\}\medskip
				{\small \it \\##1\\}
}}
{\centering
 \vskip 3em
 \large \lineskip .75em
 \@author}
 \par}}
\def\@makedate{\vskip 1.5em
 {\raggedright \small \noindent\@pubdate \par}}
\def\@makeabstract{\vskip 1.5em
{\small
\begin{center}
{\bf ABSTRACT\vspace{-.5em}\vspace{0pt}}
\end{center}
\quotation \@abstract \endquotation}}
\def\maketitle{\titlepage
\let\footnotesize\small \setcounter{page}{0}
\def\thefootnote{\arabic{footnote}}
\@makepub
\vfil
\@maketitle
\@makeauthor
\vfil
\@makeabstract
\@thanks
\vfil
\@makedate
\if@restonecol\twocolumn \else \eject \fi
\titlerelax \titleclean
\def\thefootnote{\alph{footnote}}
\setcounter{footnote}{0}
}

\ifcase\@ptsize
 \font\tenmsa=msam10
 \font\sevenmsa=msam7
 \font\fivemsa=msam5
 \font\tenmsb=msbm10
 \font\sevenmsb=msbm7
 \font\fivemsb=msbm5
\or
 \font\tenmsa=msam10 scaled \magstephalf
 \font\sevenmsa=msam8
 \font\fivemsa=msam6
 \font\tenmsb=msbm10 scaled \magstephalf
 \font\sevenmsb=msbm8
 \font\fivemsb=msbm6
\or
 \font\tenmsa=msam10 scaled \magstep1
 \font\sevenmsa=msam8
 \font\fivemsa=msam6
 \font\tenmsb=msbm10 scaled \magstep1
 \font\sevenmsb=msbm8
 \font\fivemsb=msbm6
\fi

\newfam\msafam
\newfam\msbfam
\textfont\msafam=\tenmsa  \scriptfont\msafam=\sevenmsa
  \scriptscriptfont\msafam=\fivemsa
\textfont\msbfam=\tenmsb  \scriptfont\msbfam=\sevenmsb
  \scriptscriptfont\msbfam=\fivemsb

\def\hexnumber@#1{\ifnum#1<10 \number#1\else
 \ifnum#1=10 A\else\ifnum#1=11 B\else\ifnum#1=12 C\else
 \ifnum#1=13 D\else\ifnum#1=14 E\else\ifnum#1=15 F\fi\fi\fi\fi\fi\fi\fi}

\def\msa@{\hexnumber@\msafam}
\def\msb@{\hexnumber@\msbfam}
\mathchardef\boxdot="2\msa@00
\mathchardef\boxplus="2\msa@01
\mathchardef\boxtimes="2\msa@02
\mathchardef\square="0\msa@03
\mathchardef\blacksquare="0\msa@04
\mathchardef\centerdot="2\msa@05
\mathchardef\lozenge="0\msa@06
\mathchardef\blacklozenge="0\msa@07
\mathchardef\circlearrowright="3\msa@08
\mathchardef\circlearrowleft="3\msa@09
\mathchardef\rightleftharpoons="3\msa@0A
\mathchardef\leftrightharpoons="3\msa@0B
\mathchardef\boxminus="2\msa@0C
\mathchardef\Vdash="3\msa@0D
\mathchardef\Vvdash="3\msa@0E
\mathchardef\vDash="3\msa@0F
\mathchardef\twoheadrightarrow="3\msa@10
\mathchardef\twoheadleftarrow="3\msa@11
\mathchardef\leftleftarrows="3\msa@12
\mathchardef\rightrightarrows="3\msa@13
\mathchardef\upuparrows="3\msa@14
\mathchardef\downdownarrows="3\msa@15
\mathchardef\upharpoonright="3\msa@16

\mathchardef\downharpoonright="3\msa@17
\mathchardef\upharpoonleft="3\msa@18
\mathchardef\downharpoonleft="3\msa@19
\mathchardef\rightarrowtail="3\msa@1A
\mathchardef\leftarrowtail="3\msa@1B
\mathchardef\leftrightarrows="3\msa@1C
\mathchardef\rightleftarrows="3\msa@1D
\mathchardef\Lsh="3\msa@1E
\mathchardef\Rsh="3\msa@1F
\mathchardef\rightsquigarrow="3\msa@20
\mathchardef\leftrightsquigarrow="3\msa@21
\mathchardef\looparrowleft="3\msa@22
\mathchardef\looparrowright="3\msa@23
\mathchardef\circeq="3\msa@24
\mathchardef\succsim="3\msa@25
\mathchardef\gtrsim="3\msa@26
\mathchardef\gtrapprox="3\msa@27
\mathchardef\multimap="3\msa@28
\mathchardef\therefore="3\msa@29
\mathchardef\because="3\msa@2A
\mathchardef\doteqdot="3\msa@2B

\mathchardef\triangleq="3\msa@2C
\mathchardef\precsim="3\msa@2D
\mathchardef\lesssim="3\msa@2E
\mathchardef\lessapprox="3\msa@2F
\mathchardef\eqslantless="3\msa@30
\mathchardef\eqslantgtr="3\msa@31
\mathchardef\curlyeqprec="3\msa@32
\mathchardef\curlyeqsucc="3\msa@33
\mathchardef\preccurlyeq="3\msa@34
\mathchardef\leqq="3\msa@35
\mathchardef\leqslant="3\msa@36
\mathchardef\lessgtr="3\msa@37
\mathchardef\backprime="0\msa@38
\mathchardef\risingdotseq="3\msa@3A
\mathchardef\fallingdotseq="3\msa@3B
\mathchardef\succcurlyeq="3\msa@3C
\mathchardef\geqq="3\msa@3D
\mathchardef\geqslant="3\msa@3E
\mathchardef\gtrless="3\msa@3F
\mathchardef\sqsubset="3\msa@40
\mathchardef\sqsupset="3\msa@41
\mathchardef\vartriangleright="3\msa@42
\mathchardef\vartriangleleft="3\msa@43
\mathchardef\trianglerighteq="3\msa@44
\mathchardef\trianglelefteq="3\msa@45
\mathchardef\bigstar="0\msa@46
\mathchardef\between="3\msa@47
\mathchardef\blacktriangledown="0\msa@48
\mathchardef\blacktriangleright="3\msa@49
\mathchardef\blacktriangleleft="3\msa@4A
\mathchardef\vartriangle="3\msa@4D
\mathchardef\blacktriangle="0\msa@4E
\mathchardef\triangledown="0\msa@4F
\mathchardef\eqcirc="3\msa@50
\mathchardef\lesseqgtr="3\msa@51
\mathchardef\gtreqless="3\msa@52
\mathchardef\lesseqqgtr="3\msa@53
\mathchardef\gtreqqless="3\msa@54
\mathchardef\Rrightarrow="3\msa@56
\mathchardef\Lleftarrow="3\msa@57
\mathchardef\veebar="2\msa@59
\mathchardef\barwedge="2\msa@5A
\mathchardef\doublebarwedge="2\msa@5B
\mathchardef\angle="0\msa@5C
\mathchardef\measuredangle="0\msa@5D
\mathchardef\sphericalangle="0\msa@5E
\mathchardef\varpropto="3\msa@5F
\mathchardef\smallsmile="3\msa@60
\mathchardef\smallfrown="3\msa@61
\mathchardef\Subset="3\msa@62
\mathchardef\Supset="3\msa@63
\mathchardef\Cup="2\msa@64

\mathchardef\Cap="2\msa@65

\mathchardef\curlywedge="2\msa@66
\mathchardef\curlyvee="2\msa@67
\mathchardef\leftthreetimes="2\msa@68
\mathchardef\rightthreetimes="2\msa@69
\mathchardef\subseteqq="3\msa@6A
\mathchardef\supseteqq="3\msa@6B
\mathchardef\bumpeq="3\msa@6C
\mathchardef\Bumpeq="3\msa@6D
\mathchardef\lll="3\msa@6E

\mathchardef\ggg="3\msa@6F

\mathchardef\circledS="0\msa@73
\mathchardef\pitchfork="3\msa@74
\mathchardef\dotplus="2\msa@75
\mathchardef\backsim="3\msa@76
\mathchardef\backsimeq="3\msa@77
\mathchardef\complement="0\msa@7B
\mathchardef\intercal="2\msa@7C
\mathchardef\circledcirc="2\msa@7D
\mathchardef\circledast="2\msa@7E
\mathchardef\circleddash="2\msa@7F
\def\ulcorner{\delimiter"4\msa@70\msa@70 }
\def\urcorner{\delimiter"5\msa@71\msa@71 }
\def\llcorner{\delimiter"4\msa@78\msa@78 }
\def\lrcorner{\delimiter"5\msa@79\msa@79 }
\def\yen{\mathhexbox\msa@55 }
\def\checkmark{\mathhexbox\msa@58 }
\def\circledR{\mathhexbox\msa@72 }
\def\maltese{\mathhexbox\msa@7A }
\mathchardef\lvertneqq="3\msb@00
\mathchardef\gvertneqq="3\msb@01
\mathchardef\nleq="3\msb@02
\mathchardef\ngeq="3\msb@03
\mathchardef\nless="3\msb@04
\mathchardef\ngtr="3\msb@05
\mathchardef\nprec="3\msb@06
\mathchardef\nsucc="3\msb@07
\mathchardef\lneqq="3\msb@08
\mathchardef\gneqq="3\msb@09
\mathchardef\nleqslant="3\msb@0A
\mathchardef\ngeqslant="3\msb@0B
\mathchardef\lneq="3\msb@0C
\mathchardef\gneq="3\msb@0D
\mathchardef\npreceq="3\msb@0E
\mathchardef\nsucceq="3\msb@0F
\mathchardef\precnsim="3\msb@10
\mathchardef\succnsim="3\msb@11
\mathchardef\lnsim="3\msb@12
\mathchardef\gnsim="3\msb@13
\mathchardef\nleqq="3\msb@14
\mathchardef\ngeqq="3\msb@15
\mathchardef\precneqq="3\msb@16
\mathchardef\succneqq="3\msb@17
\mathchardef\precnapprox="3\msb@18
\mathchardef\succnapprox="3\msb@19
\mathchardef\lnapprox="3\msb@1A
\mathchardef\gnapprox="3\msb@1B
\mathchardef\nsim="3\msb@1C
\mathchardef\napprox="3\msb@1D
\mathchardef\varsubsetneq="3\msb@20
\mathchardef\varsupsetneq="3\msb@21
\mathchardef\nsubseteqq="3\msb@22
\mathchardef\nsupseteqq="3\msb@23
\mathchardef\subsetneqq="3\msb@24
\mathchardef\supsetneqq="3\msb@25
\mathchardef\varsubsetneqq="3\msb@26
\mathchardef\varsupsetneqq="3\msb@27
\mathchardef\subsetneq="3\msb@28
\mathchardef\supsetneq="3\msb@29
\mathchardef\nsubseteq="3\msb@2A
\mathchardef\nsupseteq="3\msb@2B
\mathchardef\nparallel="3\msb@2C
\mathchardef\nmid="3\msb@2D
\mathchardef\nshortmid="3\msb@2E
\mathchardef\nshortparallel="3\msb@2F
\mathchardef\nvdash="3\msb@30
\mathchardef\nVdash="3\msb@31
\mathchardef\nvDash="3\msb@32
\mathchardef\nVDash="3\msb@33
\mathchardef\ntrianglerighteq="3\msb@34
\mathchardef\ntrianglelefteq="3\msb@35
\mathchardef\ntriangleleft="3\msb@36
\mathchardef\ntriangleright="3\msb@37
\mathchardef\nleftarrow="3\msb@38
\mathchardef\nrightarrow="3\msb@39
\mathchardef\nLeftarrow="3\msb@3A
\mathchardef\nRightarrow="3\msb@3B
\mathchardef\nLeftrightarrow="3\msb@3C
\mathchardef\nleftrightarrow="3\msb@3D
\mathchardef\divideontimes="2\msb@3E
\mathchardef\varnothing="0\msb@3F
\mathchardef\nexists="0\msb@40
\mathchardef\mho="0\msb@66
\mathchardef\thorn="0\msb@67
\mathchardef\beth="0\msb@69
\mathchardef\gimel="0\msb@6A
\mathchardef\daleth="0\msb@6B
\mathchardef\lessdot="3\msb@6C
\mathchardef\gtrdot="3\msb@6D
\mathchardef\ltimes="2\msb@6E
\mathchardef\rtimes="2\msb@6F
\mathchardef\shortmid="3\msb@70
\mathchardef\shortparallel="3\msb@71
\mathchardef\smallsetminus="2\msb@72
\mathchardef\thicksim="3\msb@73
\mathchardef\thickapprox="3\msb@74
\mathchardef\approxeq="3\msb@75
\mathchardef\succapprox="3\msb@76
\mathchardef\precapprox="3\msb@77
\mathchardef\curvearrowleft="3\msb@78
\mathchardef\curvearrowright="3\msb@79
\mathchardef\digamma="0\msb@7A
\mathchardef\varkappa="0\msb@7B
\mathchardef\hslash="0\msb@7D
\mathchardef\hbar="0\msb@7E
\mathchardef\backepsilon="3\msb@7F
\def\Bbb{\ifmmode\let\next\Bbb@\else
 \def\next{\errmessage{Use \string\Bbb\space only in math mode}}\fi\next}
\def\Bbb@#1{{\Bbb@@{#1}}}
\def\Bbb@@#1{\fam\msbfam#1}

\def\bk {{\hskip 0.2 cm}}

\def\acknowledgements{\@startsection{section}{4}
{\z@}{-3.5ex plus -1ex minus -.2ex}{2.3ex plus .2ex}{\normalsize\bf}
{Acknowledgements}}

\newcommand{\tab}[1]{{\sc Tab.}\,{\sf #1}}	  
\newcommand{\eq}[1]{{\sc Eq.}\,{\sf (#1)}}	  
\newcommand{\refoth}[1]{{\sf #1}}  	  	  

\def\bbbz {\Bbb{Z}}		  

\def\bbbn {\Bbb{N}}   	          
\def\bbbc {\Bbb{C}}

\newtheorem{definition}{Definition}[section]
\newtheorem{theorem}[definition]{Theorem}
\newtheorem{lemma}[definition]{Lemma}
\newtheorem{proposition}[definition]{Proposition}
\newcounter{defs}[section]

\newcommand{\be}{\begin{equation}}
\newcommand{\ee}{\end{equation}}
\newcommand{\bea}{\begin{eqnarray}}
\newcommand{\eea}{\end{eqnarray}}

\newcommand{\bdf}{\stepcounter{defs}\begin{definition}}
\newcommand{\edf}{\end{definition}}
\newcommand{\bth}{\stepcounter{defs}\begin{theorem}}
\newcommand{\eth}{\end{theorem}}
\newcommand{\blm}{\stepcounter{defs}\begin{lemma}}
\newcommand{\elm}{\end{lemma}}
\newcommand{\bpr}{\stepcounter{defs}\begin{proposition}}
\newcommand{\epr}{\end{proposition}}

\newcommand{\eprf}{\hfill $\Box$ \\}
\newcounter{pics}
\newcommand{\bpic}[4]{\begin{center}\begin{picture}(#1,#2)(#3,#4)
\refstepcounter{pics}}
\renewcommand{\thepics}{{\sf\roman{pics}}}
\newcommand{\epic}[1]{\end{picture}\\
{\small {\sc Fig.} \thepics \bk #1} \end{center}}
\newcommand{\epicspl}{\end{picture}\\		
\addtocounter{pics}{-1}\end{center}}		

\renewcommand{\thefootnote}{\rm{\alph{footnote}}}
\newcounter{tabs}
\newcommand{\btab}[1]{\refstepcounter{tabs}\begin{center}
\begin{tabular}{#1}}
\renewcommand{\thetabs}{{\sf\alph{tabs}}}
\newcommand{\etab}[1]{\end{tabular}\\[1.5ex]
{\small {\sc Tab.} \thetabs \bk #1} \end{center}}

\def\noi {\noindent}

\newcommand{\ket}[1]{\left| {#1} \right\rangle}	
\newcommand{\spn}[1]{{\rm span}\{{#1}\}}	
\newcommand{\vm}[1]{{\langle #1 \rangle}}	

\def\pmb#1{\setbox0=\hbox{#1}%
 \kern-.025em\copy0\kern-\wd0
 \kern.05em\copy0\kern-\wd0
 \kern-.025em\raise.0433em\box0 }

\def\cQ{{\cal Q}}
\def\cG{{\cal G}}
\def\cL{{\cal L}}
\def\cH{{\cal H}}

\makeatother

\def\ta{{\sf T}}       
\def\vm{{\cal V}}        
\def\vir{{\sf V}}        
\def\salg{{\cal A}}      
\def\ordering{{\cal O}}  
\def\cset{{\cal C}}      
\newcommand{\osm}[1]{{{<}_{{}_{#1}}}}           

\title{The Adapted Ordering Method for Lie Algebras and Superalgebras and
their Generalizations}
\author{Beatriz Gato-Rivera\thanks{Also known as B. Gato}
\address{Instituto de Matem\'aticas y F\'\i sica Fundamental, CSIC,\\
Serrano 123, Madrid 28006, Spain \\[.3cm]
NIKHEF-H, Kruislaan 409, NL-1098 SJ Amsterdam, The Netherlands}}
\pubdate{June 2007 \\ Revised version: November 2007}
\pubnumber{IMAFF-FM-07/01\\ NIKHEF-2007-13}

\abstract{In 1998 the Adapted Ordering Method was developed for the
representation theory of the superconformal algebras in two dimensions.
It allows: to determine maximal dimensions for a given type of space of
singular vectors, to identify all singular vectors by only a few coefficients,
to spot subsingular vectors and to set the basis for constructing embedding
diagrams. In this article we present the Adapted Ordering Method for general 
Lie algebras and superalgebras, and their generalizations,  provided they 
can be triangulated.  We also review briefly the results obtained for the 
Virasoro algebra and for the $N=2$ and Ramond $N=1$ superconformal algebras.}

\begin{document}

\maketitle



\section{Introduction and Notation}

In 1998 the Adapted Ordering Method was developed by M. D\"{o}rrzapf 
and B. Gato-Rivera\cite{SD1}, for the study of the representation theory of the 
superconformal algebras in two dimensions, also known as super Virasoro algebras.
These are infinite-dimensional Lie superalgebras which contain the Virasoro
algebra as a subalgebra. They were first constructed three decades ago 
independently by Kac, along with his classification of Lie superalgebras\cite{Kac0},
and by Ademollo et al. as the symmetry algebras of the supersymmetric 
strings\cite{ademollo}. At present, although several research lines make use 
of the superconformal algebras, their main relevance in physics is still the fact
that they provide the underlying symmetries of Superstring Theory. The
superconformal symmetries have a number $N$ of fermionic anticommuting
currents, corresponding to $N$ supersymmetries. Their mode decomposition
provide the $N$ infinite sets of anticommuting generators of the superconformal
algebras, whereas the Virasoro operators provide the infinite set of commuting
generators, together with some other infinite sets of commuting generators
which exist for $N>1$ and arise as symmetries between the supercurrents.
The Adapted Ordering Method was applied successfully to the $N=2$ superconformal 
algebras\cite{SD1,SD2} (topological, Neveu-Schwarz, Ramond and twisted) 
and to the Ramond $N=1$ superconformal algebra\cite{Ramond}, allowing to 
obtain rigorous proofs for several conjectured results, as well as many new 
results, especially for the case of the twisted $N=2$ superconformal 
algebra and the case of the Ramond $N=1$ superconformal algebra.

An obvious question now is whether the Adapted Ordering Method can be
generalized and can be applied to the study of algebras different than the 
superconformal ones.  The answer is positive and the purpose of this article is 
precisely to provide the general description of the Adapted Ordering Method for  
Lie algebras and superalgebras, and their generalizations, provided they  
have a triangular decomposition, as is the case for many of them\cite{Moody}. 

Let us introduce some basic concepts and notation which will be used throughout 
this article. For a given algebra or superalgebra one defines freely generated 
modules over a highest weight (h.w.) vector, denoted as {\it Verma modules}. The
{\it annihilation operators} of the algebra are the generators which annihilate
the h.w.  vectors of the Verma modules, whereas the {\it creation operators} 
are the generators directly involved in the construction of the Verma modules
by acting on the h.w. vectors.
A Verma module is in general irreducible, but in some degenerate cases
it contains submodules which are freely generated over, at least, one h.w. 
vector different from the h.w. vector of the Verma module. 
These vectors are annihilated by all the annihilation operators of the algebra,
consequently, and are usually referred to as {\it singular vectors}. 
The irreducible h.w. representations are then obtained as the 
quotients of the Verma modules divided by all their submodules. Surprisingly, 
the complete set of singular vectors do not generate all the submodules 
in the case of Verma modules which contain {\it subsingular vectors}. The 
reason is that subsingular vectors are singular vectors of the quotient space, 
but not of the Verma module itself\cite{subsing,beatriz1,DGR1,beatriz2}. In this 
case one has to divide further by the submodules generated by the subsingular 
vectors, repeating this division procedure successively, if necessary. 

On the Verma modules one introduces a hermitian contravariant form,
known as Shapovalov form. 
The vanishing of the corresponding determinant indicates the existence of 
at least one singular vector. The determinant may not detect the whole set
of singular vectors, however, neither does it give the dimension of the space 
of singular vectors with some given weights. There could be in fact more than
one linearly independent singular vectors with the same weights. Therefore,
the dimensions of the spaces of singular vectors have to be found by an
independent procedure. The Adapted Ordering Method provides such a procedure
since it puts upper limits on these dimensions, allowing to determine the
maximal dimension for a given space of singular vectors.  For most weights 
of a Verma module these upper limits on the dimensions of the spaces
of singular vectors are found equal to zero and, as a consequence, one obtains
a rigorous proof that there cannot exist any singular vectors for these weights. 
For some weights, however, one finds that spaces of singular vectors are
allowed to exist, either only one-dimensional, as is the case for the Virasoro 
algebra, or even higher dimensional spaces, as it happens for the $N=2$ and 
Ramond $N=1$ superconformal algebras\cite{SD2,Ramond,beatriz2,thesis,paper2}. 
As we will see, the Adapted Ordering Method also allows to identify all singular 
vectors by only a few coefficients, to spot subsingular vectors and to set the 
basis for constructing embedding diagrams, as a result. 

The idea for developing  the Adapted Ordering Method originated, 
in rudimentary form, from
a procedure due to A. Kent for the study of the representations of the
Virasoro algebra\cite{adrian1}. For this purpose the author analytically
continued the Virasoro Verma modules, yielding `generalised' Verma
modules, where he constructed `generalised' singular vectors 
in terms of analytically continued Virasoro operators. This analytical 
continuation is not necessary, however, for the Adapted Ordering Method,
nor is it necessary to construct singular vectors in order to apply it. The
underlying idea is the concept of {\it adapted orderings} for all the possible
terms of the `would be' singular vectors. An adapted ordering is a criterion,
satisfying certain requirements, to decide which of two given terms is the
bigger one. To be more specific, a total ordering will be called {\it adapted}
to a subset of terms provided some conditions are met. The complement of 
that subset will be the {\it ordering kernel} and will play a crucial r\^ole since
its size puts un upper limit on the dimension of the space of singular vectors.  

In what follows, in section 2 we will describe the Adapted Ordering Method for 
a general Lie algebra or superalgebra with a triangular decomposition and, as an 
example, we will apply this method to the Virasoro algebra. In section 3 we will 
review briefly the results obtained for the $N=2$ and the Ramond $N=1$ 
superconformal algebras, as an illustration of the 
possibilities of this method. Section 4 is devoted to conclusions.


\section{The Adapted Ordering Method}
\label{sec:AOM}

Let $\salg$ denote a Lie algebra or superalgebra with a triangular 
decomposition: $\salg = \salg^- \oplus {\cal H}_{\salg} \oplus \salg^+ $,
where $\salg^-$ is the set of {\it creation operators},  $\salg^+$ is the set 
of {\it annihilation operators}, and ${\cal H}_{\salg}$ is the {\it Cartan 
subalgebra}.  In general, an eigenvector with respect to the Cartan 
subalgebra with {\it relative weights} given by the set $\{l_i \}$, in particular 
a singular vector $\Psi_{\{l_i\}}$, can be expressed as a sum of products 
of creation operators with total weights $\{l_i\}$ acting on a  h.w. vector with
weights $\{\Delta_i\} $: 
\bea
\Psi_{\{l_i \}}&=& \sum_{m_1,m_2,....\in\bbbn_0}^{ }
\sum_{a,b,c,...} k_{a_{-1}^{m_1},a_{-2}^{m_2},...b_{-1}^{n_1},b_{-2}^{n_2},.....} \, 
X_ {\{l_i\}}^{a_{-1}^{m_1},a_{-2}^{m_2},...b_{-1}^{n_1},b_{-2}^{n_2},.....}
\ket{\{\Delta_i\}} \,, \label{eq:psil}
\eea
where $a_{-1}, a_{-2},..... b_{-1}, b_{-2},.....$ are the creation operators of the
algebra, $X_ {\{l_i\}}^{a_{-1}^{m_1},a_{-2}^{m_2},...b_{-1}^{n_1},b_{-2}^{n_2},.....} $ are 
the products of the creation operators: $a_{-1}^{m_1}  a_{-2}^{m_2} 
..... b_{-1}^{n_1}  b_{-2}^{n_2}.....$,
with total weights $\{l_i\}$, which will be denoted simply as {\it terms},
and $k_{a_{-1}^{m_1},a_{-2}^{m_2},...b_{-1}^{n_1},b_{-2}^{n_2},.....} \in\bbbc$ 
are coefficients which depend on the given term. A non-trivial term Y then refers
to a term with non-trivial coefficient $k_Y$. Observe that the weights of $\Psi_{\{l_i \}}$
are given by $\{l_i+ \Delta_i\}$, it is however customary to label the
vectors in the Verma modules by their {\it relative weights} $\{l_i\}$.

Now let us define the set $\cset_{\{l_i\}}$  as the set of all the terms with 
weights $\{l_i\}$:
\bea
\cset_{\{l_i\}} &=
& \{X_ {\{l_i\}}^{a_{-1}^{m_1},a_{-2}^{m_2},...b_{-1}^{n_1},b_{-2}^{n_2},.....} , 
\, m_1, m_2,.... n_1, n_2,.....\in\bbbn_0 \} \,,
\eea
and let $\ordering$ denote a total ordering on $\cset_{\{l_i\}}$, 
that is an ordering such that any two different terms in 
$\cset_{\{l_i\}}$ are ordered with respect to each other.
Thus $\Psi_{\{l_i\}}$ in \eq{\ref{eq:psil}} needs to contain an
$\ordering$-{\it smallest} $X_0\in\cset_{\{l_i\}}$ with $k_{X_0}\neq 0$ and 
$k_Y=0$ for all $Y\in\cset_{\{l_i\}}$ with $Y\osm{\ordering} X_0$ and 
$Y\neq X_0$. We define an {\it adapted ordering} on $\cset_{\{l_i\}}$ as
follows:

\bdf \label{def:adapt}
A total ordering $\ordering$ on $\cset_{\{l_i\}}$  is called 
adapted to the subset $\cset^{A}_{\{l_i\}}\subset\cset_{\{l_i\}}$ in the Verma 
module $\vm_{\{\Delta_i \}}$ if for any element $X_0\in\cset^{A}_{\{l_i\}}$ at 
least one annihilation operator $\Gamma $ exists for which
$\Gamma \, X_0 \ket{\{\Delta_i \}}$ contains a non-trivial term $\tilde{X}$
\bea
\Gamma \, X_0 \ket{\{\Delta_i \}} &=&
( k_{\tilde{X}} \tilde{X} + ....... ) \, \ket{\{\Delta_i \}}
\label{eq:adapt1}
\eea
which is absent, however, for all $\Gamma \, X \ket{\{\Delta_i \}}$,
where $X$ is any term $X \in\cset_{\{l_i\}}$ which is $\ordering$-{\it larger}
than $X_0$, that is such that
$X_0 \osm{\ordering} X$. The complement of $\cset^{A}_{\{l_i\}}$,
${\ }\cset^{K}_{\{l_i\}}=\cset_{\{l_i\}} \setminus \cset^{A}_{\{l_i\}}$ 
is the kernel with respect to the ordering $\ordering$ in the Verma 
module $\vm_{\{\Delta_i \}}$.
\edf

Now we will see that the coefficients with respect to the terms of the ordering 
kernel $\cset^{K}_{\{l_i\}}$ uniquely identify a singular vector $\Psi_{\{l_i \}}$. 
Since the size of the ordering kernels are in general small, 
it turns out that just a few coefficients completely determine
a singular vector no matter its size, what allows to find easily product
expressions for descendant singular vectors. For example, in the case
of the conformal and N=1,2 superconformal algebras the ordering kernels
found for most weights have zero or one term, for some weights they have 
two terms and for some other weights they have three terms.
This property is summarized in the following theorem:

\bth \label{th:kernel}
Let $\ordering$ denote an ordering adapted to $\cset^{A}_{\{l_i\}}$ at 
weights $\{l_i\}$ with kernel $\cset^{K}_{\{l_i\}}$ for a given Verma module 
$\vm_{\{\Delta_i \}}$. If two singular vectors $\Psi^{1}_{\{l_i\}}$ and 
$\Psi^{2}_{\{l_i\}}$ with the same weights have $k_{X}^1=k_{X}^2$ 
for all $X\in\cset^{K}_{\{l_i\}}$, then
\bea
\Psi^{1}_{\{l_i\}} &\equiv & \Psi^{2}_{\{l_i\}}\,.
\eea
\eth

{\bf Proof of Theorem}  \refoth{\ref{th:kernel}}: 
Let us consider the singular vector ${\Psi}_{\{l_i\}} = \Psi^{1}_{\{l_i\}} - \Psi^{2}_{\{l_i\}}$, 
which does not contain any terms of the ordering kernel $\cset^K_{\{l_i\}}$, simply 
because $k_{X}^1=k_{X}^2$ for all $X\in\cset^{K}_{\{l_i\}}$. As $\cset_{\{l_i\}}$
is a totally ordered set with respect to $\ordering$,
the non-trivial terms of ${\Psi}_{\{l_i\}}$, provided ${\Psi}_{\{l_i\}}$ is non-trivial,
need to have a $\ordering$-smallest $X_0\in\cset^{A}_{\{l_i\}}$. Thus the 
coefficient $k_{X_0}$ of $X_0$ in ${\Psi}_{\{l_i\}}$ must be non-trivial.  As 
$\ordering$ is adapted to $\cset^{A}_{\{l_i\}}$ one can find an annihilation 
operator $\Gamma $ such that $\Gamma X_0\ket{\{\Delta_i\}}$ contains a 
non-trivial term that cannot 
be created by $\Gamma $ acting on any other term of ${\Psi}_{\{l_i\}}$ which 
is $\ordering$-larger than $X_0$. But $X_0$ was chosen to be the
$\ordering$-smallest term of ${\Psi}_{\{l_i\}}$. Therefore,
$\Gamma X_0\ket{\{\Delta_i\}}$ contains a non-trivial term that cannot be 
created from any other term of ${\Psi}_{\{l_i\}}$. The coefficient of this term 
is obviously given by $ck_{X_0}$ with $c$ a non-trivial complex number. But 
${\Psi}_{\{l_i\}}$ is a singular vector and therefore must be annihilated by any 
annihilation operator, in particular by $\Gamma$. It follows that $k_{X_0}=0$,
contrary to our original assumption. Thus, the set of non-trivial terms of 
${\Psi}_{\{l_i\}}$ is empty and therefore ${\Psi}_{\{l_i\}}=0$. This results in
$\Psi^{1}_{\{l_i\}} = \Psi^{2}_{\{l_i\}}$.
\eprf

Theorem \refoth{\ref{th:kernel}} states, therefore, that if two singular vectors 
with the same weights, in the same Verma module, agree on the coefficients
of the ordering kernel, then they are identical. A crucial point now is that the size
of the kernel puts an upper limit on the dimension of the corresponding 
space of singular vectors, as stated in the following theorem:

\bth \label{th:dims}
Let $\ordering$ denote an ordering adapted to $\cset^{A}_{\{l_i\}}$ at 
weights $\{l_i\}$ with kernel $\cset^{K}_{\{l_i\}}$ for a given Verma module 
$\vm_{\{\Delta_i \}}$. If the ordering kernel $\cset^{K}_{\{l_i\}}$ has $n$ 
elements, then there are at most $n$ linearly independent singular vectors
$\Psi_{\{l_i\}}$ in $\vm_{\{\Delta_i \}}$ with relative weights $\{l_i\}$.
\eth

{\bf Proof of Theorem} \refoth{\ref{th:dims}}:
 
Suppose there were more than $n$ linearly independent singular vectors
$\Psi_{\{l_i\}}$ in $\vm_{\{\Delta_i\}}$ with relative weights ${\{l_i\}}$. We 
choose $n+1$ linearly independent singular vectors among them 
$\Psi_1$,$\ldots$,$\Psi_{n+1}$. The ordering kernel $\cset^{K}_{\{l_i\}}$ 
has the $n$ elements $X_1$,$\ldots$,$X_n$. Let $k_{jk}$ denote the 
coefficient of the term $X_j$ in the vector $\Psi_k$ in a suitable basis
decomposition. The coefficients $k_{jk}$ thus form a $n$ by $n+1$ matrix 
$M$. The homogeneous system of linear equations $M\lambda=0$ thus has 
a non-trivial solution $\lambda^0=(\lambda^0_1,\ldots,\lambda^0_{n+1})$
for the vector $\lambda$. We then form the linear combination 
$\Psi=\sum_{i=1}^{n+1} \lambda^0_i\Psi_i$. Obviously, the coefficient of $X_j$ 
for the vector $\Psi$  is just given by the $j$-th component of the vector 
$M\lambda$ which is trivial for $j=1,\ldots,n$. Hence, the coefficients of $\Psi$
are trivial on the ordering kernel. On the other hand, $\Psi$ is a linear
combination of singular vectors and therefore it is also a singular vector.
Due to theorem \refoth{\ref{th:kernel}} one immediately finds that 
$\Psi\equiv 0$ and therefore $\sum_{i=1}^{n+1}\lambda_i\Psi_i=0$. This, 
however, contradicts the assumption that $\Psi_1$,$\ldots$, $\Psi_{n+1}$ are 
linearly independent.
\eprf

Therefore, one needs to find suitable orderings in
order to obtain the smallest possible kernels. 
Observe that the maximal possible dimension $n$ does not imply that
all the singular vectors of the corresponding type are $n$-dimensional.
From this theorem one deduces that if $\cset^{K}_{\{l_i\}}=\emptyset$ for
a given Verma module, then there are no singular vectors with relative weights
$\{l_i\}$ in it. That is:

\bth \label{th:striv}
Let $\ordering$ denote an ordering adapted to $\cset^{A}_{\{l_i\}}$ at 
weights $\{l_i\}$ with trivial kernel $\cset^{K}_{\{l_i\}}=\emptyset$ for a given 
Verma module $\vm_{\{\Delta_i \}}$. A singular vector $\Psi_{\{l_i\}}$ 
in $\vm_{\{\Delta_i \}}$ with relative weights $\{l_i\}$ must be therefore trivial. 
\eth

Although this theorem is deduced straightforwardly from theorem 
\refoth{\ref{th:dims}}, which is exactly proven, there is another interesting 
proof using theorem \refoth{\ref{th:kernel}}

{\bf Proof of Theorem} \refoth{\ref{th:striv}}: The trivial vector $0$ satisfies 
any annihilation conditions for any weights. As the ordering kernel is trivial, 
$\cset^{K}_{\{l_i\}}=\emptyset$, the components of the vectors $0$ and $\Psi_{\{l_i\}}$
agree on the ordering kernel and using theorem \refoth{\ref{th:kernel}}
we obtain $\Psi_{\{l_i\}}=0$.
\eprf

\vskip .4cm

As a simple example of the Adapted Ordering Method we will see now the  
application of this method to the Virasoro algebra $\vir$, which has been
extensively studied in the literature\cite{Kac,FeFu,RCW}. 
This algebra is 
given by the commutation relations
\bea
\left[L_m,L_n\right] = (m-n) L_{m+n} +\frac{C}{12} (m^3-m) \delta_{m+n,0}
\,, & \left[C,L_m\right] = 0 \,, & m,n\in\bbbz \,,
\eea
where $C$ commutes with all operators of $\vir$ and can hence be taken to be 
constant $c\in\bbbc$. $\vir$ can be written in its {\it triangular decomposition}:
$\vir=\vir^-\oplus {\cal H}_{\vir} \oplus \vir^+$, where $\vir^-=\spn{L_{-m}:m\in\bbbn}$ 
is the set of {\it creation operators}, $\vir^+=\spn{L_m:m\in{\bf {N}}}$ 
is the set of {\it annihilation operators}, and the {\it Cartan subalgebra} is given by 
${\cal H}_{\vir}=\spn{L_0,C}$. For elements of $\vir$ that are eigenvectors of $L_0$ 
with respect to the adjoint representation the $L_0$-eigenvalue is usually 
called the {\it level} $l$.
The {\it terms} are obviously given by the products of the 
form $L_{-p_I}\ldots L_{-p_1}$, $p_q\in\bbbn$ for $q=1,\ldots, I$, 
$I\in\bbbn$, with level $l=\sum_{q=1}^{I}p_q$. Note that annihilation 
operators $L_m \in \vir^+$ have negative level $l= -m$, $m\in{\bf {N}}$.

A representation with $L_0$-eigenvalues bounded from below contains a 
highest weight (h.w.) vector $\ket{\Delta}$, with $L_0$-eigenvalue $\Delta$, 
which is annihilated by the set of annihilation operators $\vir^+$ :
\bea
\vir^+ \ket{\Delta}=0\,, & L_0\ket{\Delta}=\Delta\ket{\Delta}\,. \label{eq:hwc}
\eea
The Verma module $\vm_{\Delta}$ built on $\ket{\Delta}$
is $L_0$-graded in a natural way. The corresponding $L_0$-eigenvalue 
is called the {\it conformal weight} and is written for convenience as $\Delta+l$, 
where $l$ is the {\it level}. Any proper submodule of $\vm_{\Delta}$ needs to 
contain a {\it singular vector} $\Psi_l$ that is not proportional to the h.w. vector 
$\ket{\Delta}$ but still satisfies the  h.w. conditions with conformal weight $\Delta+l$: 
\bea
\vir^+\Psi_l=0 \,, & L_0\Psi_l=(\Delta+l)\Psi_l \,. \label{eq:hwc2}
\eea

Now we will see the total ordering on the set of terms 
$\cset_l$ at level $l$ defined by Kent\cite{adrian1} 
for the Virasoro algebra. One has to take into account, however, that Kent used 
the following ordering in order to show that, in his generalised Verma modules,
the generalised singular vectors at  level $0$ 
satisfying the h.w. conditions are actually proportional 
to the h.w. vector. Using the Adapted Ordering technology, though, one 
deduces that this ordering already implies that all Virasoro singular vectors are 
unique at their levels up to proportionality, simply because the ordering kernel 
for each level $l\in\bbbn$ has just one element: $L_{-1}^l$.

\bdf \label{def:virorder}
On the set $\cset_l$ of terms of Virasoro operators at level l 
one introduces the total ordering $\ordering_{\vir}$ for $l\in\bbbn$:
For any two terms $X_1, X_2\in\cset_l$, 
$X_1\neq X_2$, with $X_i=L_{-m^i_{I_i}}\ldots L_{-m^i_1}L_{-1}^{n^i}$,
$n^i=l-m^i_{I_i}\ldots-m^i_1$, or $X_i=L_{-1}^l$, $i=1,2$ one defines
\bea
X_1 \osm{\ordering_{\vir}}X_2 & {\rm if} & n^1>n^2\,. \label{eq:virord1}
\eea
If, however, $n^1=n^2$ one computes the index
$j_0=\min\{j:m^1_j-m^2_j\neq 0,j=1,\ldots,\min(I_1,I_2)\}$. One then defines
\bea
X_1 \osm{\ordering_{\vir}}X_2 & {\rm if} & m^1_{j_0}<m^2_{j_0}\,.
\eea
For $X_1=X_2$ one sets $\, X_1 \osm{\ordering_{\vir}}X_2$ and
$\, X_2\osm{\ordering_{\vir}}X_1$.
\edf
The index $j_0$ describes the first mode, read from the right to the left, for 
which the generators in $X_1$ and $X_2$ ($L_{-1}$ excluded) are different. 
For example, in $\cset_8$ one has  $L_{-2}L_{-2}L_{-2}L_{-1}^2 
\osm{\ordering_{\vir}} L_{-4}L_{-2}L_{-1}^2$ with index $j_0=2$. Observe that 
$L_{-1}^l\in\cset_l$ is the $\ordering_{\vir}$-smallest term in $\cset_l$. Now 
using the Adapted Ordering Method one finds the following theorem\cite{SD1}. 

\bth \label{th:viradap}
The ordering $\ordering_{\vir}$ is adapted to $\cset^A_l=
\cset_l\setminus\{L_{-1}^l\}$ for each level $l\in\bbbn$ and
for all Verma modules $\vm_{\Delta}$. The ordering kernel is
given by the single element set $\cset^K_l=\{L_{-1}^l\}$.
\eth
For example let us consider the set of terms at level 3, 
$\cset_3=\{L^3_{-1}, L_{-2}L_{-1}, L_{-3}\}$. 
One finds the total ordering 
$L^3_{-1} \osm{\ordering_{\vir}} L_{-2}L_{-1} \osm{\ordering_{\vir}} L_{-3}$,
which is adapted to $\cset^A_3=\{L_{-2}L_{-1}, L_{-3}\}$ with the ordering
kernel $\cset^K_3=\{L_{-1}^3\}$. To see this one has to compute 
the action of the annihilation operators $\Gamma \in \{L_1, L_2, L_3\}$ 
on the three terms. In fact, the action of $L_1$ already reveals the
structure of $\cset^A_3$, as $L_1 L_{-2} L_{-1} \ket{\Delta}$ contains 
the term $L^2_{-1}$ that is absent in $L_1 L_{-3} \ket{\Delta}$. The
action of the three annihilation operators on $L^3_{-1} \ket{\Delta}$,
however, produce terms that are also created by the action of these
operators on $L_{-2} L_{-1} \ket{\Delta}$ and/or $L_{-3} \ket{\Delta}$.

Finally, from the previous theorem one now deduces the known result about
the uniqueness of the Virasoro singular vectors\cite{adrian1}.

\bth \label{th:viruni}
If the Virasoro Verma module $\vm_{\Delta}$ contains a singular vector 
$\Psi_l$ at level $l$, $l\in\bbbn$, then $\Psi_l$ is unique up to proportionality. 
\eth


\section{Results for the superconformal algebras}
\label{sec:Nis2}

As an illustration of the possibilities of the Adapted Ordering Method, in this section we
will review briefly the results obtained for the $N=2$ and Ramond $N=1$ superconformal 
algebras. This method has been applied to the topological, to the Neveu-Schwarz
and to the Ramond  $N=2$ algebras in Ref. \icite{SD1}, to the twisted $N=2$ algebra 
in Ref. \icite{SD2} and to the Ramond $N=1$ algebra in Ref. \icite{Ramond}. 
As the representation theory of these superconformal algebras
has different types of Verma modules, one has to introduce different adapted 
orderings for each type and the corresponding kernels also allow different 
degrees of freedom. 

Let us start with the topological $N=2$ superconformal algebra $\ta$. 
It contains the Virasoro generators $\cL_m$ with trivial central extension,
a Heisenberg algebra $\cH_m$ corresponding to a U(1) current, and the
fermionic generators $\cG_m$ and $\cQ_m$ corresponding to
two anticommuting fields with conformal weights 2 and 1 respectively.
$\ta$ satisifies the (anti-)commutation relations\cite{DVV}

\be
\begin{array}{ll}
\left[\cL_m,\cL_n\right] = (m-n)\cL_{m+n}\,, &
\left[\cH_m,\cH_n\right] = \frac{C}{3}m\delta_{m+n}\,,\\
\left[\cL_m,\cG_n\right] = (m-n)\cG_{m+n}\,, &
\left[\cH_m,\cG_n\right] = \cG_{m+n}\,,\\
\left[\cL_m,\cQ_n\right] = -n\cQ_{m+n}\,, &
\left[\cH_m,\cQ_n\right] = -\cQ_{m+n}\,,\\
\left[\cL_m,\cH_n\right] = -n\cH_{m+n}+\frac{C}{6}(m^2+m)\delta_{m+n}\,,\\
\left\{\cG_m,\cQ_n\right\} = 2\cL_{m+n}-2n\cH_{m+n}
+\frac{C}{3}(m^2+m)\delta_{m+n}\,,\\
\left\{\cG_m,\cG_n\right\} = \left\{\cQ_m,\cQ_n\right\} = 0\,, &
 m,~n\in\bbbz\,.\label{topalgebra}
\end{array}
\ee 
The set of {\it annihilation operators} which is common for all the Verma
modules, $\ta^+$, is spanned by the generators with positive index, the set of 
{\it creation operators} which is common for all the Verma modules, $\ta^-$, is 
spanned by the generators with negative index, and the {\it zero modes} are 
given by $\ta^0=\spn{\cL_0,\cH_0,C,\cG_0,\cQ_0}$. The Cartan subalgebra 
is generated by ${\cal H}_{\ta}=\spn{\cL_0,\cH_0, C}$, where $C$ can 
be taken to be constant $c\in\bbbc$, and the fermionic generators 
$\{\cG_0,\cQ_0\}$ can act as annihilation or as creation operators, classifying
the different types of Verma modules in this way.

A h.w. vector $\ket{\Delta,h}^{\cal N}$ is an eigenvector of ${\cal H}_{\ta}$ 
with $\cL_0$-eigenvalue $\Delta$, $\cH_0$-eigenvalue $h$, and vanishing
$\ta^{+}$ action. Additional vanishing conditions ${\cal N}$ are possible with 
respect to the operators $\cG_0$ and $\cQ_0$, resulting as 
follows\cite{beatriz2}. One can distinguish four different types of h.w.
vectors $\ket{\Delta,h}^{\cal N}$ labeled by a superscript
${\cal N}\in\{G,Q,GQ\}$, or no superscript at all: h.w. vectors 
$\ket{\Delta,h}^{G}$ annihilated by $\cG_0$ but not by $\cQ_0$  
($\cG_0$-closed), h.w. vectors $\ket{\Delta,h}^{Q}$ annihilated 
by $\cQ_0$ but not by $\cG_0$ ($\cQ_0$-closed), h.w. vectors
$\ket{0,h}^{GQ}$ annihilated by both $\cG_0$ and $\cQ_0$ ({\it chiral}), 
with zero conformal weight necessarily, and finally undecomposable 
h.w. vectors  $\ket{0,h}$ that are neither annihilated by 
$\cG_0$ nor by $\cQ_0$  ({\it no-label}), also with zero conformal weight.
Hence we have four different types of Verma modules\cite{beatriz2}:
$\vm_{\Delta,h}^G$, $\vm_{\Delta,h}^Q$, $\vm_{0,h}^{GQ}$ and 
$\vm_{0,h}$, built on the four different types of h.w. vectors. 

For elements $X$ of $\ta$ which are eigenvectors of ${\cal H}_{\ta}$ with 
respect to the adjoint representation one defines the {\it level} $l$ as 
$[\cL_0,X]= l X$ and the {\it charge} $q$ as $[\cH_0,X]= q X$. In particular, 
elements of the form
$X=\cL_{-l_L} \ldots \cL_{-l_1}\cH_{-h_H} \ldots \cH_{-h_1}
\cQ_{-q_Q} \ldots \cQ_{-q_1}\cG_{-g_G} \ldots \cG_{-g_1}$,
and any reorderings of it, have level $l=\sum_{j=1}^{L}l_j
+\sum_{j=1}^{H}h_j+\sum_{j=1}^{Q}q_j+\sum_{j=1}^{G}g_j$ and
charge $q=G-Q$. The Verma modules are naturally
$\bbbn_0\times\bbbz$ graded with respect to the ${\cal H}_{\ta}$
eigenvalues relative to the eigenvalues $(\Delta,h)$ of the h.w. vector. For a 
vector labeled as $\Psi_{l,q}$ in $\vm_{\Delta,h}^N$ the $\cL_0$-eigenvalue 
is $\Delta+l$ and the $\cH_0$-eigenvalue is $h+q$ with the level 
$l\in\bbbn_0$ and the relative charge $q\in\bbbz$. 

The singular vectors are annihilated by $\ta^+$ and may also satisfy additional 
vanishing conditions with respect to the operators $\cG_0$ and $\cQ_0$. 
Therefore one also distinguishes singular vectors of the types\cite{beatriz2} 
$\Psi^G_{l,q}$, $\Psi^Q_{l,q}$, $\Psi^{GQ}_{l,q}$ and $\Psi_{l,q}$. 
As there are $4$ types of Verma modules and $4$ types of singular vectors 
one might think of $16$ different combinations of singular vectors in Verma 
modules. However, no-label and chiral singular vectors do not exist neither in 
{\it chiral} Verma modules $\vm_{0,h}^{GQ}$ nor in {\it no-label} Verma 
modules\cite{beatriz2} $\vm_{0,h}$ (with one exception: 
chiral singular vectors exist at level 0 in no-label Verma modules).
Using the Adapted Ordering Method one has to introduce adapted orderings
for the remaining $12$ combinations, whose kernels give upper limits for
the dimensions of the corresponding spaces of singular vectors. One finds that 
for most charges $q$ singular vectors do not exist. For the case of the
Verma modules $\vm_{\Delta,h}^G$ built on $\cG_0$-closed
h.w. vectors $\ket{\Delta,h}^G$, for $c\neq 3$, the maximal
dimensions for the spaces of singular vectors $\Psi_{l,q}^{G}$, $\Psi_{l,q}^{Q}$
$\Psi_{l,q}^{GQ}$ and $\Psi_{l,q}$ are given as follows\cite{SD1}:

\btab{|l|c|c|c|c|c|}
\hline \label{tab:dim1}
 & $q=-2$ & $q=-1$ & $q=0$ & $q=1$ & $q=2$ \\
\hline
$\Psi_{l,q,\ket{\Delta,h}^G}^{G}$ &
$0$ & $1$ & $2$ & $1$ & $0$ \\
\hline
$\Psi_{l,q,\ket{\Delta,h}^G}^{Q}$ &
$1$ & $2$ & $1$ & $0$ & $0$ \\
\hline
$\Psi_{l,q,\ket{-l,h}^G}^{GQ}$ &
$0$ & $1$ & $1$ & $0$ & $0$ \\
\hline
$\Psi_{l,q,\ket{-l,h}^G}$ &
$0$ & $1$ & $1$ & $0$ & $0$ \\
\hline
\etab{Maximal dimensions for singular vectors spaces in $\vm_{\Delta,h}^G$.}
Charges $q$ that are not given have dimension $0$ and hence do not allow 
any singular vectors. The maximal dimensions for the case of the Verma 
modules $\vm_{\Delta,h}^Q$, for $c\neq 3$, are obtained simply by 
interchanging $G \leftrightarrow Q$ and $q \leftrightarrow -q$ in the previous 
table.

For the case of singular vectors in chiral Verma modules $\vm_{0,h}^{GQ}$ 
and in no-label Verma modules $\vm_{0,h}$, for $c\neq 3$, one obtains 
the following maximal dimensions\cite{SD1}:

\btab{|l|c|c|c|c|c|}
\hline \label{tab:dim3}
 & $q=-2$ & $q=-1$ & $q=0$ & $q=1$ & $q=2$ \\
\hline
$\Psi_{l,q,\ket{0,h}^{GQ}}^{G}$ &
$0$ & $0$ & $1$ & $1$ & $0$ \\
\hline
$\Psi_{l,q,\ket{0,h}^{GQ}}^{Q}$ &
$0$ & $1$ & $1$ & $0$ & $0$ \\
\hline
$\Psi_{l,q,\ket{0,h}}^{G}$ &
$0$ & $1$ & $3$ & $3$ & $1$ \\
\hline
$\Psi_{l,q,\ket{0,h}}^{Q}$ &
$1$ & $3$ & $3$ & $1$ & $0$ \\
\hline
\etab{Maximal dimensions for singular vectors spaces
in $\vm_{0,h}^{GQ}$ and in $\vm_{0,h}$.}

Tables \tab{\ref{tab:dim1}} and \tab{\ref{tab:dim3}} prove the conjecture 
made in Ref. \icite{beatriz2}, using the algebraic mechanism denoted 
{\it the cascade effect}, about the possible existing types of topological
singular vectors.  In addition, low level examples were 
constructed\cite{beatriz2} for all these types, what proves that all 
of them exist already at level 1. The four types of two-dimensional 
spaces of singular vectors of \tab{\ref{tab:dim1}} also exist starting at level 2, 
and four examples at level 3 were constructed\cite{beatriz2} as well.
For the case of the three-dimensional spaces of singular vectors in no-label 
Verma modules in \tab{\ref{tab:dim3}}, the corresponding types of singular 
vectors have been constructed at level 1 generating 
one-dimensional\cite{beatriz2} as well as two-dimensional\cite{SD1} spaces,
but no further search has been done for the three-dimensional spaces.

Transferring the dimensions we have found in tables \tab{\ref{tab:dim1}} 
and \tab{\ref{tab:dim3}} to the Neveu-Schwarz $N=2$ 
algebra\cite{ademollo,BFK,Dobrev2,Matsuo,Kiritsis} is
straightforward as this algebra is related to the topological $N=2$ 
algebra through the topological twists ${\ }T_W^{\pm}$:
${\ }\cL_m=L_m\pm 1/2 H_m$, ${\ }\cH_m=\pm H_m$,
${\ }\cG_m=G^{\pm}_{m+1/2}$ and ${\ }\cQ_m=G^{\mp}_{m-1/2}$, 
where $G^{\pm}_{m+1/2}$ are the half-integer moded fermionic 
generators. As a result, the standard Neveu-Schwarz h.w. vectors
correspond to $\cG_0$-closed topological h.w. vectors, whereas the 
chiral (antichiral) Neveu-Schwarz h.w. vectors, annihilated by $G^+_{-1/2}$ 
($G^-_{-1/2}$), correspond to chiral topological h.w. vectors.
This implies\cite{beatriz1,beatriz2} that the standard and chiral and antichiral 
Neveu-Schwarz singular vectors correspond to topological singular vectors 
of the types $\Psi^{G}_{l,q,\ket{\Delta,h}^G}$ 
and $\Psi^{GQ}_{l,q,\ket{\Delta,h}^G}$, whereas the 
Neveu-Schwarz singular vectors built in chiral 
or antichiral Verma modules correspond to topological singular 
vectors of only the type $\Psi^{G}_{l,q,\ket{\Delta,h}^{GQ}}$.
As a consequence, by untwisting the first row of table \tab{\ref{tab:dim1}} 
one recovers the results\cite{thesis,paper2} that in Verma modules of the 
Neveu-Schwarz $N=2$ algebra singular vectors only exist for charges 
$q=0, \pm 1$ and two-dimensional spaces only exist for uncharged singular
vectors. By untwisting the third row of table \tab{\ref{tab:dim1}} one gets
a proof for the conjecture\cite{beatriz2} that chiral singular vectors in 
Neveu-Schwarz Verma modules only exist for $q=0, 1$ whereas antichiral 
singular vectors only exist for $q=0, -1$. The untwisting of the first row
of table \tab{\ref{tab:dim3}}, finally, proves the 
conjecture\cite{beatriz1,beatriz2}
that in chiral Neveu-Schwarz Verma modules $\,\vm_{h/2,h}^{NS,ch}\,$ 
singular vectors only exist for $q=0, -1$, whereas in antichiral Verma 
modules $\,\vm_{-h/2,h}^{NS,a}\,$ singular vectors only exist for $q=0, 1$.

As to the representations of the Ramond $N=2$ 
algebra\cite{BFK,Dobrev2,Matsuo,Kiritsis}, they 
are exactly isomorphic to the representations of the topological $N=2$ 
algebra. Namely, combining the topological twists ${\ }T_W^{\pm}$ and the 
spectral flows one constructs a one-to-one mapping between the Ramond 
singular vectors and the topological singular vectors, at the same levels 
and with the same charges\cite{DGR3}. Therefore the results of tables
 \tab{\ref{tab:dim1}}  and \tab{\ref{tab:dim3}} can be transferred 
to the Ramond singular vectors simply by exchanging the labels
$G \to (+), {\ }Q \to (-)$, where the helicity $(\pm)$ denotes the vectors 
annihilated by the fermionic zero modes $G_0^{\pm}$, and by taking
into account that the chiral and undecomposable {\it no-helicity} Ramond
vectors\cite{beatriz2,DGR1,DGR3}, require conformal weight 
$\Delta+l=c/24 \,$. 

The twisted $N=2$ superconformal algebra\cite{BFK,Dobrev2,Matsuo,Kiritsis}
is not related to the other three $N=2$ algebras. It has mixed modes, integer 
and half-integer, for the fermionic generators, and the eigenvectors have
no charge, as the U(1) current generators are half-moded, but they have
{\it fermionic parity}. The Adapted Ordering Method was worked out for
the twisted $N=2$ algebra in Ref. \icite{SD2}. The maximal dimension for 
the spaces of singular vectors in standard Verma modules was found to be two 
and these two-dimensional singular spaces were shown to exist by explicit 
computation, starting at level $3/2$. In Verma modules built on $G_0$-closed
h.w. vectors, however, the singular vectors were found to be only
one-dimensional. This method also allowed to propose a reliable conjecture
for the coefficients of the relevant terms of all singular vectors, i.e. for
the coefficients with respect to the ordering kernels, what made possible
to identify all the cases of two-dimensional spaces of singular vectors for all
levels, as well as to identify all $G_0$-closed singular vectors.
The resulting expressions, in turn, led to the discovery of 
subsingular vectors for this algebra, and several explicit examples were
also computed. Finally, the multiplication rules for singular vectors
operators were derived using the ordering kernel coefficients, what set
the basis for the analysis of the twisted $N=2$ embedding diagrams.    

Finally let us consider the $N=1$ superconformal
algebras\cite{Kac,Kiritsis,MRC,Dobrev1}. 
The structure of the h.w. representations of the Neveu-Schwarz $N=1$ 
algebra has been completely understood in Ref. \icite{Astash}. The 
corresponding Verma modules do not contain two-dimensional singular vector 
spaces neither subsingular vectors. In the case of the Ramond $N=1$ algebra,
however, the application of the Adapted Ordering Method in Ref. \icite{Ramond}
has shown that its representations have a much richer structure than previously
suggested in the literature. In particular, it was found that standard Verma modules
may contain two-dimensional spaces of singular vectors and also subsingular vectors.
Moreover, the two-dimensional ordering kernels allowed to derive multiplication rules 
for singular vector operators and led to expressions for the two-dimensional spaces of 
singular vectors. Using these multiplication rules descendant singular vectors were
studied and embedding diagrams were derived for the rational models. In addition,
this allowed to conjecture the ordering kernel coefficients of all singular vectors
and therefore identify these vectors uniquely.   

\section{Conclusions and Final Remarks}

We have presented the Adapted Ordering Method for general Lie algebras and 
superalgebras, and their generalizations, provided they can be triangulated, 
as is the case in many interesting examples. This method 
is based on the concept of adapted orderings, 
leading to the definition of the ordering kernels, which play a crucial 
r\^ole since their sizes limit the dimensions of the corresponding spaces of
singular vectors. As a result the adapted orderings must be chosen such that 
the ordering kernels are as small as possible. Weights for which the ordering
kernels are trivial do not allow any singular vectors in the corresponding weight
spaces. On the other hand, non-trivial ordering kernels give us the maximal
dimensions of the possible spaces of singular vectors and uniquely define all 
singular vectors through the coefficients with respect to them, allowing to set
the basis for constructing embedding diagrams.

The Adapted Ordering Method follows from the Definition 2.A 
plus the Theorems 2.B, 2.C and 2.D, which are rigorously proven. 
There is nothing in the Definition 2.A, neither in the three theorems, that 
restricts the application of this method to infinite-dimensional algebras.
For the same reason, it seems clear that  the Adapted Ordering Method 
should be useful also for generalized Lie algebras and superalgebras such 
as affine Kac-Moody algebras, non-linear W-algebras, superconformal 
W-algebras, loop Lie algebras, Borcherds algebras, F-Lie algebras for $F>2$ 
($F=1$ are Lie algebras and $F=2$ are Lie superalgebras), etc. 

One may wonder whether there are any prescriptions in order to construct
the most suitable orderings with the smallest kernels.The answer to this question 
is that there are no general prescriptions or recipes as the orderings depend entirely
on the given algebras. The way to proceed is a matter of trial and error. That is,
one constructs one total ordering first, that can always be done since a total ordering
is simply a definition establishing which of two given terms is the bigger one, 
then one computes the kernel and decides whether this
kernel is small enough. In the case it is not, then one constructs a second
ordering and repeats the procedure until one finds a suitable ordering. It may
also happen, for a given algebra, that this procedure does not give any useful
information because all the total orderings one can construct are adapted only to
the empty subset, in which case the ordering kernel is the whole set of terms:
$\cset^{K}_{\{l_i\}} = \cset_{\{l_i\}}$.

The Adapted Ordering 
Method has been applied so far to the $N=2$ and Ramond $N=1$ 
superconformal algebras, allowing to prove several conjectured results as
well as to obtain many new results, as we have reviewed. For example,
this method allowed to discover subsingular vectors and two-dimensional 
spaces of singular vectors for the twisted $N=2$ and Ramond $N=1$ 
algebras\cite{SD2,Ramond}. (For the other three isomorphic $N=2$ algebras
two-dimensional singular spaces had been discovered\cite{beatriz2,thesis,paper2},
as well as subsingular vectors\cite{subsing,beatriz1,DGR1,beatriz2}, before the
Adapted Ordering Method was applied to them). We are convinced therefore that
this method should be of very much help for the study of the representation
theory of many other algebras, in particular the $N>2$ superconformal algebras,
and some (at least) of the generalized Lie algebras listed above.

\vskip .5in
\acknowledgements

I am grateful to Christoph Schweigert for providing some information about
triangulated algebras and to Bert Schellekens for reading carefully the
manuscript.
The work of the author is partially supported by funding of the spanish
Ministerio de Educaci\'on y Ciencia, Research Project FPA2005-05046.

\noi



\end{document}